\UseRawInputEncoding 
\documentclass[superscriptaddress,aps,prd,reprint,onecolumn,amsmath,amssymb,showpacs,nofootinbib]{revtex4-2}
\usepackage{graphicx}%
\usepackage{multirow}%
\usepackage{amsmath,amssymb,amsfonts}%
\usepackage{amsthm}%
\usepackage{mathrsfs}%
\usepackage[title]{appendix}%
\usepackage{xcolor}%
\usepackage{textcomp}%
\usepackage{booktabs}%
\usepackage{algorithm}%
\usepackage{algorithmicx}%
\usepackage{algpseudocode}%
\usepackage{listings}%
\usepackage{etoolbox}%
\usepackage{float}
\usepackage[hidelinks]{hyperref}
\usepackage{textcomp}

\theoremstyle{thmstyleone}%

\theoremstyle{thmstyletwo}%

\theoremstyle{thmstylethree}%

\newcommand{\beq}{\begin{equation}}
\newcommand{\eeq}{\end{equation}}
\newcommand{\bea}{\begin{eqnarray}}
\newcommand{\eea}{\end{eqnarray}}

\newcommand{\gsim}{\lower.7ex\hbox{$\;\stackrel{\textstyle>}{\sim}\;$}}
\newcommand{\lsim}{\lower.7ex\hbox{$\;\stackrel{\textstyle<}{\sim}\;$}}

\newcommand{\be}{\begin{equation}}
\newcommand{\ee}{\end{equation}}
\newcommand{\ba}{\begin{eqnarray}}
\newcommand{\ea}{\end{eqnarray}}

\def \d {{\rm d}}
\makeatletter

\newcommand{\bmhead}[1]{%
  \par\medskip\noindent\textbf{#1}\enspace\ignorespaces
}

\makeatother

\makeatletter
\renewcommand{\bibliography}[1]{}
\@ifundefined{bibliography*}{}{\renewcommand{\bibliography*}[1]{}}
\makeatother

\begin{document}

\title{Inference on inner galaxy structure via gravitational waves from supermassive binaries}

\author{Yifan Chen}
\email{chen.yifan@sjtu.edu.cn}
\affiliation{State Key Laboratory of Dark Matter Physics, Tsung-Dao Lee Institute, Shanghai Jiao Tong University, Shanghai 200240, China}
\affiliation{Key Laboratory for Particle Astrophysics and Cosmology (MOE) \& Shanghai Key Laboratory for Particle Physics and Cosmology, Shanghai Jiao Tong University, Shanghai 200240, China}
\affiliation{Center of Gravity, Niels Bohr Institute, Blegdamsvej 17, 2100 Copenhagen, Denmark}

\author{Matthias Daniel}
\affiliation{Institute for Theoretical Physics, Goethe University, 60438 Frankfurt am Main, Germany}

\author{Daniel J.\ D'Orazio}
\affiliation{Space Telescope Science Institute, 3700 San Martin Drive, Baltimore, MD 21218, USA}
\affiliation{Department of Physics and Astronomy, Johns Hopkins University, 3400 North Charles Street, Baltimore, Maryland 21218, USA}
\affiliation{Niels Bohr International Academy, Niels Bohr Institute, Blegdamsvej 17, 2100 Copenhagen, Denmark}

\author{Xuanye Fan}
\email{xuanye.fan@ligo.org}
\affiliation{Department of Physics and Astronomy, Stony Brook University, Stony Brook, NY 11794, USA}

\author{Andrea Mitridate}
\affiliation{Deutsches Elektronen-Synchrotron DESY, Notkestr. 85, 22607 Hamburg, Germany}

\author{Laura Sagunski}
\affiliation{Institute for Theoretical Physics, Goethe University, 60438 Frankfurt am Main, Germany}

\author{Xiao Xue}
\email{xxue@ifae.es}
\affiliation{Institut de Física d'Altes Energies (IFAE), The Barcelona Institute of Science and Technology, Campus UAB, 08193 Bellaterra (Barcelona), Spain}
\affiliation{II. Institute of Theoretical Physics, Universit\"{a}t Hamburg, 22761 Hamburg, Germany}
\affiliation{Deutsches Elektronen-Synchrotron DESY, Notkestr. 85, 22607 Hamburg, Germany}

\author{The NANOGrav Collaboration (A list of consortium authors and their affiliations appears at the end of the paper.)}

\begin{abstract}
The detection of a stochastic gravitational wave background by pulsar-timing arrays indicates the presence of a population of supermassive black hole binaries. Although the observed spectrum generally matches predictions for orbital evolution driven by gravitational-wave emission in circular orbits, there is a preference for a spectral turnover at the lowest observed frequencies, which may point to substantial hardening during a transition from early environmental influences to later stages dominated by emission. In the vicinity of these binaries, the ejection of stars or dark matter particles through gravitational three-body slingshots efficiently extracts orbital energy, leading to a low-frequency turnover in the spectrum. Here we model how the gravitational-wave spectrum depends on the initial inner galactic profile before scouring by binary ejections while accounting for a range of initial binary eccentricities. By analysing the NANOGrav 15-year data, we find that a parsec-scale galactic-centre density of around $10^6 M_{\odot} \mathrm{pc}^{-3}$ is favoured across most of the parameter space, thus shedding light on the environmental effects that shape black hole evolution and the combined matter density near galaxy centres.
\end{abstract}

\maketitle

\section*{Introduction}\label{sec1}

Recent advances of pulsar-timing arrays (PTAs), which can leverage precise measurements of timing residuals within a galactic-scale detector, have ushered in a new era of stochastic gravitational wave background (SGWB) detection. The SGWB, defined by a superposition of incoherent gravitational waves (GWs), initially emerged as a common-spectrum process~\cite{Arzoumanian:2020vkk}. Subsequent data provided evidence of a quadrupolar correlation function~\cite{NANOGrav:2023gor,EPTA:2023fyk,Reardon:2023gzh,Xu:2023wog}, famously known as the Hellings-Downs curve~\cite{Hellings:1983fr}, which affirmed the presence and characteristics of the SGWB.

The observed spectrum of the SGWB is consistent with expectations for a population of supermassive black hole binaries (SMBHBs), that is dominated by binaries with comparable mass ratios, total masses in the range $10^{9.2-10.4}\,M_{\odot}$, and redshifts from $0.15$ to $0.9$~\cite{NANOGrav:2023hfp,EPTA:2023xxk}, where $M_{\odot}$ represents the solar mass. Although the spectrum is consistent with a steady slope due to GW emission by circular binaries, the NANOGrav 15-year data~\cite{NANOGrav:2023hfp} exhibit a mild preference for a low-frequency turnover, a feature also indicated by data from the Parkes Pulsar Timing Array~\cite{Reardon:2023gzh} and European Pulsar Timing Array~\cite{EPTA:2023xxk}, which have over 10 years of observations. This feature implies an acceleration in the rate of orbital hardening, which would be a potential solution to the final parsec problem~\cite{Begelman:1980vb,Milosavljevic:2002ht}. 

A natural and simple explanation involves environmental interactions with stars and cold dark matter, which are noted for their potentially high density in galactic centres~\cite{Navarro:1995iw,Gondolo:1999ef,Genzel:2010zy}. Both stars and dark matter can be expelled from a system through gravitational slingshots during encounters with binary components, thereby extracting orbital energy~\cite{Quinlan:1996vp}. This process involves three-body scattering, for which the energy extraction efficiency is substantially higher than that of two-body dynamical friction~\cite{Chandrasekhar:1943ys}, especially when the binary components have comparable masses and are sufficiently close. Such three-body slingshot interactions can substantially alter the density profile of the galactic centre, particularly flattening the inner distribution~\cite{Milosavljevic:2001vi,Merritt:2002jz}. This underscores the importance of considering the co-evolution of the density profile and the binary orbit. A pivotal study by Sesana and Khan~\cite{Sesana:2015haa} demonstrates that the orbital hardening rate observed in $N$-body simulations can be effectively approximated by results from scattering simulations~\cite{Quinlan:1996vp} within environments characterized by the distribution at the SMBHB influence radius~\cite{Frank:1976uy} before scouring. In this study, we use the NANOGrav $15$-year dataset to investigate how the SGWB is shaped by initial galactic-centre density profiles and binary orbital eccentricities. By jointly modeling the co-evolution of eccentricity and environmental hardening through three-body scattering, we demonstrate that current PTA data, despite their marginal sensitivity, already encode measurable information about parsec-scale matter distributions in galactic centres.

\section*{Binary Hardening by Three-body Scattering}

The SGWB emanating from SMBHBs is an incoherent superposition of signals from individual sources. Its spectrum is characterized by the characteristic strain $h_c(f)$, obtained by integrating the single-binary GW emission over redshift $z$, total mass $M$ and mass ratio $q \leq 1$ and weighted by the comoving population distribution ${\rm d}^3\eta/({\rm d}z\,{\rm d}M\,{\rm d}q)$, where $\eta$ is the co-moving volumetric number density of binaries. The emitted GW frequency in the source frame, $f_s$ is related to the observed frequency by the usual redshift relation $f = f_s/(1+z)$.

The single-binary GW emission spectrum is inversely proportional to the orbital hardening rate ${\rm d}a/{\rm d}t$, as binaries that spend longer at a given orbital separation contribute more strongly to the SGWB. The semimajor axis $a$ satisfies Kepler's law, $a^3 = GM/(2\pi f_{\rm orb})^2$, where $G$ is Newton's constant and the orbital frequency $f_{\rm orb} = f_s/2$ for circular binaries. The total hardening rate is ${\rm d}a/{\rm d}t \equiv \sum_j ({\rm d}a/{\rm d}t)|_j$, with the index $j$ accounting for all relevant mechanisms driving orbital decay.

For circular binaries evolving purely under GW emission, the semimajor axis is
\be
\frac{{\rm d}a}{{\rm d}t}\Big|_{\rm GW}^{e=0}
= -\frac{64}{5} \frac{G^3 M^3}{c^5 a^3} \frac{q}{(1+q)^2},
\label{eq:hardenGW}
\ee
where $c$ is the speed of light and $e$ is the eccentricity. This scaling leads directly to the well-known power-law spectrum of the SGWB, $h_c \propto f^{-2/3}$, for a population of circular, GW-driven binaries. Eccentric orbits enhance GW emission over a range of frequencies, as discussed in Methods. Consequently, a turnover in the spectrum at lower frequencies is possible before the orbit undergoes circularization through GW emission~\cite{Enoki:2006kj}.

In galactic centres, SMBHBs form within a background of stars and particle dark matter, once their separation falls below the influence radius $r_i$. This radius is defined as the distance at which the total enclosed mass of stars and dark matter is twice the mass of the SMBHB~\cite{Frank:1976uy}. Given the extreme mass ratio between the SMBHB and the stars and dark matter, both act effectively as test particles. Each may undergo several gravitational encounters with one of the black holes (BHs) until it gains sufficient kinetic energy to be ejected. This three-body slingshot process becomes efficient as the semimajor axis $a$ approaches the hardening radius, defined as $a_h \equiv r_i q/(4(1+q)^2)$~\cite{Quinlan:1996vp}. The rate of orbital hardening due to three-body scattering, averaged over a background of particles with matter density $\rho$ and velocity dispersion $\sigma$, is given by~\cite{Quinlan:1996vp}:
\be \frac{\d a}{\d t}\Big{|}_{\text{3BS}} = -H G\frac{\rho}{\sigma} a^2,\label{eq:hardenSC}\ee
where $H$ is a dimensionless coefficient typically ranging from $15$ to $20$, as observed in scattering simulations~\cite{Quinlan:1996vp}. It is important to note that three-body scattering is fundamentally distinct from two-body dynamical friction~\cite{Chandrasekhar:1943ys} and becomes more dominant within the hardening radius. Assuming a constant $\rho/\sigma$ ratio over time, the spectral evolution for circular orbits follows $h_c \propto f$. Note that three-body scatterings tend to increase the eccentricity on average~\cite{Quinlan:1996vp}, in contrast to the effects of GW emission.

Historically, the three-body slingshot process was thought to stall orbital evolution by ejecting background stars and depleting the loss cone, a region of phase space consisting of low-angular-momentum orbits in which stars are close enough for strong interactions with the binary. This led to what is known as the final parsec problem~\cite{Begelman:1980vb}. However, $N$-body simulations~\cite{Khan:2011gi,Preto:2011gu,Khan:2012fp} have shown that merger-induced triaxiality can efficiently repopulate the loss cone~\cite{Merritt:2003pf}.

These simulations also reveal that two initially cuspy profiles, each centred on a BH with peak densities, will merge and flatten into a single core profile following the merger~\cite{Milosavljevic:2001vi,Merritt:2002jz}. This underscores the need for comprehensive simulations that simultaneously address the co-evolution of the SMBHB orbit and the density profile. Further analysis comparing results from scattering simulations~\cite{Quinlan:1996vp} with $N$-body simulation outcomes~\cite{Khan:2012fp} demonstrates that predictions of orbital evolution can closely align with equation~(\ref{eq:hardenSC}), assuming that $\rho/\sigma$ remains constant, as determined by the initial profile at the influence radius $r_i$ of the SMBHB~\cite{Sesana:2015haa}. This statement is supported by observations that the total mass ejected during SMBHB evolution is approximately of the order of $M$~\cite{Quinlan:1996vp,Merritt:1999ry,Milosavljevic:2001vi,Antonini:2015cqa,Antonini:2015sza,Celoria:2018mzr} and is primarily distributed within the influence radius at the onset. During simulations, the loss cone at this radius remains fully populated as it is driven by the efficient diffusion of particles in triaxial potentials.

\begin{figure}[H]
    \centering
   \includegraphics[width=0.5\textwidth]{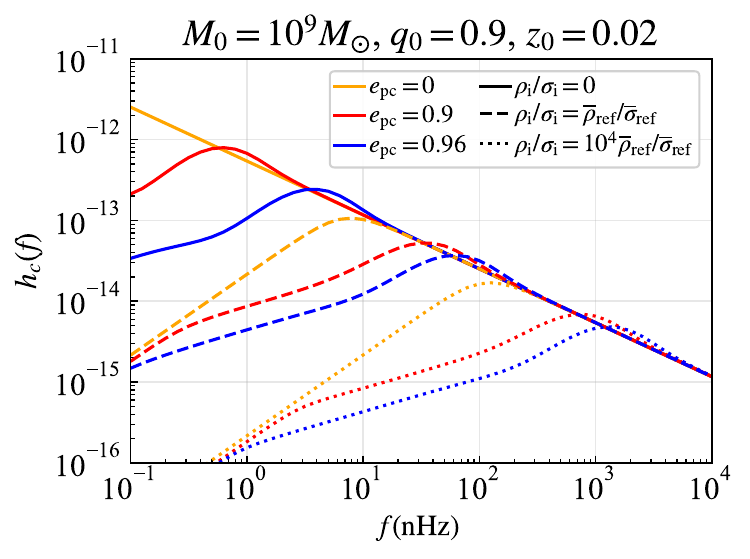}
   \caption{\textbf{Illustrative SGWB spectra from a simplified SMBHB population.} 
Examples of the SGWB spectra derived from an SMBHB population modelled as delta functions in the space of $(M,q,z)$, specifically $\mathrm{d}^3\eta/(\mathrm{d}z\,\mathrm{d}M\,\mathrm{d}q) = \delta(M-M_0)\delta(z-z_0)\delta(q-q_0)\,\mathrm{Mpc^{-3}}$. The values of the constants $(M_0, z_0, q_0)$ are indicated at the top. In all scenarios, the evolution of the semimajor axis begins at $1$\,pc, with the eccentricity $e_{\rm pc}$ defined at that separation. Colours indicate different $e_{\rm pc}$ values, and line styles denote variations in $\rho_i/\sigma_i$ for three-body scattering, using $\bar{\rho}_{\rm ref}\equiv 10^5\,M_\odot/\mathrm{pc}^3$ and $\bar{\sigma}_{\rm ref}\equiv 10^{-3}\,c$.}
    \label{fig:GWsingleSMBH}
\end{figure}

By comparing equation~(\ref{eq:hardenGW}) with equation~(\ref{eq:hardenSC}), one sees that the GW spectrum naturally separates into two regimes: a low-frequency phase dominated by three-body scattering and a high-frequency phase governed primarily by GW emission. The transition between these phases is marked by a turnover frequency in the source frame~\cite{Chen:2016zyo}: 
\begin{equation}\begin{split}
f_t &= \pi^{-1} \left( \frac{5c^5}{64} \frac{(1+q)^2}{q} \frac{\rho_i}{\sigma_i} H \right)^{3/10} G^{-1/10} M^{-2/5}\\
&\approx 2.8\,\text{nHz} \, \left( \frac{(1+{q})^2}{{q}}   \frac{{\rho}_i/\bar{\rho}_{\rm ref}}{{\sigma}_i/\bar{\sigma}_{\rm ref}} \frac{H}{18} \right)^{3/10} \left(\frac{M}{10^{10} M_\odot} \right)^{-2/5},
\end{split}\label{eq:ft}
\end{equation}
Here, we introduce $\rho_i$ and $\sigma_i$ as the values of $\rho$ and $\sigma$ evaluated at $r_i$, and define $\bar{\rho}_{\rm ref}\equiv 10^5\,M_\odot/\text{pc}^3 \approx 3.8\times 10^6\,c^{-2} \,\text{GeV}/\text{cm}^3$ and $\bar{\sigma}_{\rm ref} \equiv 10^{-3}c$. Figure~\ref{fig:GWsingleSMBH} presents examples of the GW spectra from an individual SMBHB across various parameter settings, demonstrating how the magnitude of three-body scattering and the initial binary eccentricity influence the spectral shape.

\section*{Inference of galactic-centre matter distributions}

The NANOGrav $15$-year data indicate a slight preference for a turnover at low frequencies, particularly around $4$\,nHz~\cite{NANOGrav:2023hfp}, 
suggesting deviations from the expected behavior of purely circular binaries driven by GW emission. Given that both eccentricity and star- and dark matter-induced three-body scatterings can contribute to this turnover, we conduct a comprehensive survey of their joint parameter space.

We adopt a straightforward power-law distribution for the galactic-centre density profile before scouring, parameterized as follows:
\be \rho(r) = \rho_{\rm pc} \left(\frac{r}{1\,\text{pc}}\right)^{-\gamma},\label{eq:DMrho}\ee
where $r$ is the radial distance, the reference radius $1$\,pc is set for our analysis, $\rho_{\mathrm{pc}}$ denotes the matter density normalization to be constrained and $\gamma$ represents the radial slope of the profile. We explore values of $\gamma$ ranging from $0$ to $2.4$, consistent with the inner region of the Dehnen density-profile family~\cite{1993MNRAS.265..250D}. For each scenario, we first determine the influence radius $r_i$ by satisfying the condition $\textstyle\int_0^{r_i}\! \rho\, 4\pi r^2 dr = 2M$~\cite{Frank:1976uy} and then calculate $\rho_i \equiv \rho(r_i)$ using equation~(\ref{eq:DMrho}) and $\sigma_i \equiv \sigma(r_i) = \sqrt{GM/r_i}$ via the virial theorem.

Our analysis targets include the initial eccentricity $e_0$, defined as occurring when the binary is formed at $a = r_i$, and the density-profile parameters $\rho_{\rm pc}$ and $\gamma$. The distribution of SMBHB parameters, namely total mass $M$, mass ratio $q$, and redshift $z$, follows the fiducial population model derived from astrophysical interpretations of NANOGrav data (the fiducial `Phenom+Astro' model without phenomenological environmental parameters) using \texttt{holodeck}~\cite{NANOGrav:2023hfp}, as detailed in Supplementary Information. The dominant contributions to the SGWB are expected from binaries with $\log_{10} (M/M_{\odot})  \in (9.2, 10.4)$, $q \in (0.5, 0.8)$, and $z \in (0.15, 0.9)$~\cite{NANOGrav:2023hfp}.

For each parameter combination of $(e_0, \rho_{\rm pc}, \gamma)$ and $(M, q, z)$, we compute the orbital and eccentricity evolution of the SMBHB, taking into account both GW emission and three-body scattering, as detailed in Methods. We then derive the total SGWB spectrum by integrating over $(M, q, z)$. Finally, we assess the likelihood that the SGWB spectrum produced by each parameter set $(e_0, \rho_{\rm pc}, \gamma)$ matches the observed spectrum in the lowest five frequency bins ($2-10$\,nHz) of the NANOGrav 15-year data, which exhibit robust signal-to-noise ratios~\cite{NANOGrav:2023gor}. We treat the overall normalization of the SMBHB distribution as a nuisance parameter, assigning it astrophysical priors consistent with how the fiducial population model was derived~\cite{NANOGrav:2023hfp}, as detailed in Supplementary Information.

In Fig.~\ref{fig:constraint}, we present the posterior distribution of the parameters $(e_0, \log_{10}(\rho_{\rm pc} [M_\odot/\text{pc}^3]/10^5), \gamma)$. The results reveal that the $1\,\sigma$ regions (dark blue) indicate the presence of three-body scatterings, with $\log_{10}(\rho_{\rm pc} [M_\odot/\text{pc}^3]/10^5)$ estimated at $0.6^{+1.4}_{-2.6}$. There is an expected degeneracy between $e_0$ and $\rho_{\rm pc}$~\cite{Taylor:2016ftv,Chen:2018znx,EPTA:2023xxk,Bi:2023tib}, where a higher $e_0$ corresponds to a lower required density. However, the $1\,\sigma$ region indicates that when $\rho_{\rm pc}$ drops below $10\,M_\odot/\text{pc}^3$, GW emission requires an extremely high initial eccentricity, $e_0 > 0.999$, to account for the turnover. This is because lower densities result in a larger $r_i$, and GW emission tends to circularize the orbit before it reaches the observed frequency range. The light blue regions represent the $2\,\sigma$ confidence interval. The white region is excluded at the $95\%$-confidence level, thereby setting an upper limit on $\rho_{\rm pc}$ as it would result in a turnover frequency inconsistent with the observational data. 

\begin{figure}[H]
    \centering    
    \includegraphics[width=0.5\textwidth]{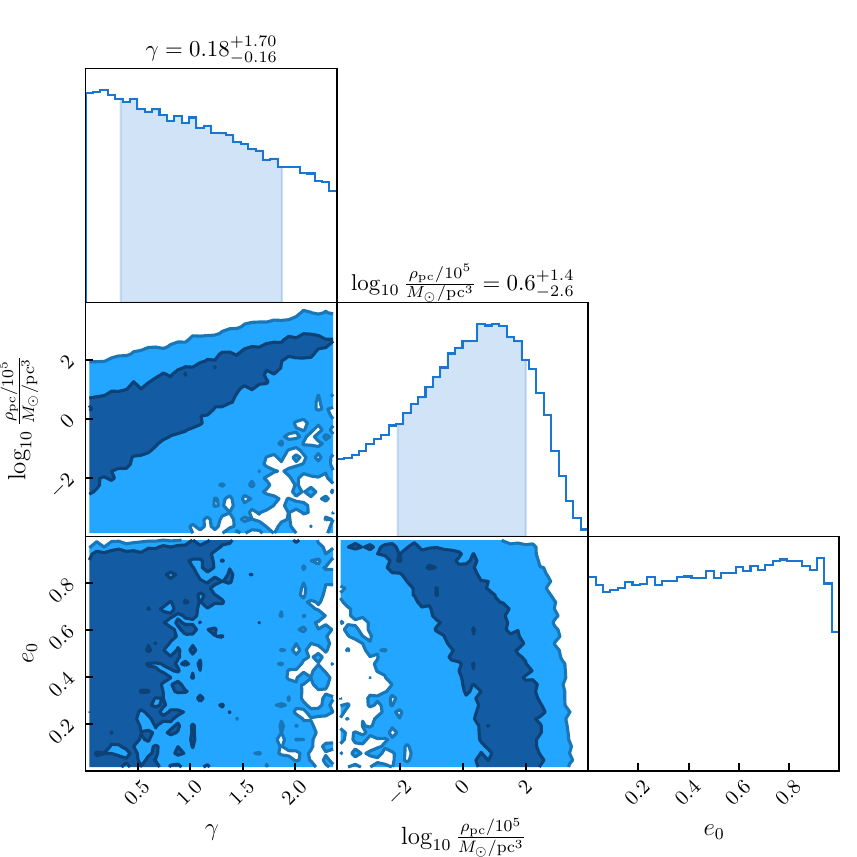}
    \caption{\textbf{Posterior constraints on eccentricity and environmental parameters.}
Posterior distribution of the parameters $(e_0, \log_{10}(\rho_{\rm pc}[M_\odot/\mathrm{pc}^3]/10^5), \gamma)$, representing the initial eccentricity and the pre-scouring density profile, inferred from the lowest five frequency bins of the NANOGrav 15-year dataset. The $1\sigma$ and $2\sigma$ confidence regions are shown in dark blue and light blue, respectively.}
    \label{fig:constraint}
\end{figure}

\begin{figure}[H]
    \centering
    \includegraphics[width=0.9\textwidth]{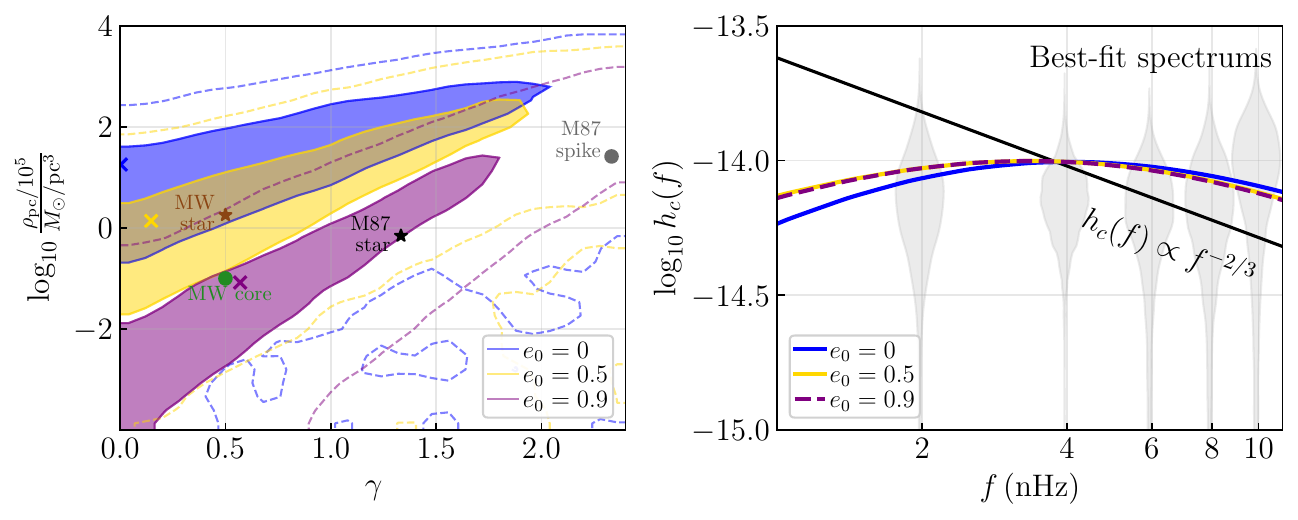}
    \caption{\textbf{Constraints and corresponding SGWB spectra for fixed initial eccentricities.}
\textbf{Left:} Posterior distribution of the density-profile parameters $(\log_{10}(\rho_{\rm pc}[M_\odot/\mathrm{pc}^3]/10^5), \gamma)$ for three representative initial eccentricities, $e_0 = 0$ (blue), $0.5$ (yellow), and $0.9$ (purple). The $1\sigma$ and $2\sigma$ confidence regions are shown in darker shades and with dashed contours, respectively. Benchmark matter-density profiles, including the stellar distribution in M87 (black star), the stellar core of the Milky Way (brown star), a dark-matter spike in M87 (gray dot), and a flattened dark-matter core in the Milky Way (green dot), are included for comparison, with parameters detailed in the text.
\textbf{Right:} Best-fit SGWB spectra corresponding to parameter choices $(\log_{10}(\rho_{\rm pc}[M_\odot/\mathrm{pc}^3]/10^5), \gamma) = (1.3,0.08)$, $(-0.09,0.06)$, and $(-1.9,0.27)$ for $e_0 = 0$, $0.5$, and $0.9$, respectively, as indicated by the crosses $(\times)$ in the left panel. The spectra are compared against the lowest five frequency bins of the NANOGrav 15-year dataset~\cite{NANOGrav:2023gor}.}
    \label{fig:PTADMC}
\end{figure}

In the left panel of Fig.~\ref{fig:PTADMC}, we present the posterior distribution for the density profile parameters ($\rho_{\rm pc}, \gamma$) for specific initial eccentricities of $e_0 = 0, 0.5,$ and $0.9$. The contours generally follow approximately constant values of $\rho_i/\sigma_i$, as suggested by equation~(\ref{eq:ft}). The right panel of Fig.~\ref{fig:PTADMC} presents the best-fit spectra for various values of $e_0$. Distributions with smaller $\gamma$ values are preferred over steeper ones because larger $\gamma$ leads to a broader range of $\rho_i/\sigma_i$ across the SMBHB population parameters $(M, q)$, which in turn results in a wider distribution of the turnover frequency $f_t$ as defined in equation~(\ref{eq:ft}). This causes the spectrum to have a broader intermediate region, requiring a normalization factor higher than the fiducial value, making high-$\gamma$ cases less favored. A conservative upper limit on $\rho_{\rm pc}(\gamma)$ is established based on the $95\%$ exclusion for $e_0 = 0$, since higher $e_0$ values lead to more stringent constraints.

For comparative purposes, we examine various benchmark star and dark matter profiles: the modelled stellar distribution in the nearby galaxy M87 with $\gamma=4/3$ and $\rho_{\rm pc} \approx 0.7\times 10^5\,M_\odot/\text{pc}^3$~\cite{McLaughlin:1998sb} (black star); the Milky Way's (MW) modelled core star distribution with $\gamma=1/2$ and $\rho_{\rm pc} \approx 1.8\times 10^5\,M_\odot/\text{pc}^3$~\cite{Merritt:2009mr} (brown star); a hypothetical dark matter spike in M87 with $\gamma = 7/3$ and $\rho_{\rm pc} \approx 2.6 \times 10^6\,M_\odot/\text{pc}^3$~\cite{Gondolo:1999ef}, formed by an adiabatically growing central SMBH from an initial Navarro-Frenk-White distribution~\cite{Navarro:1995iw} (gray dot); and a hypothetical flattened dark matter spike in the MW with $\gamma=1/2$ and $\rho_{\rm pc} \approx 10^4\,M_\odot/\text{pc}^3$, formed as the BH grew from a low-mass seed~\cite{Ullio:2001fb} (green dot). Interestingly, dark matter spikes are not favored in the hardening process because higher $\gamma$ values result in larger $r_i$ above $1$\,pc, leading to lower $\rho_i$. Conversely, core-like stellar or dark matter profiles with $\gamma = 1/2$ in the MW, as well as the stellar distribution in M87, fall naturally within the best-fit region of our analysis. Lower $\gamma$ values are expected due to flattening by previous SMBHB mergers or reformation after a galaxy merger~\cite{Ullio:2001fb}.

\begin{figure}[H]
    \centering
    \includegraphics[width=0.9\textwidth]{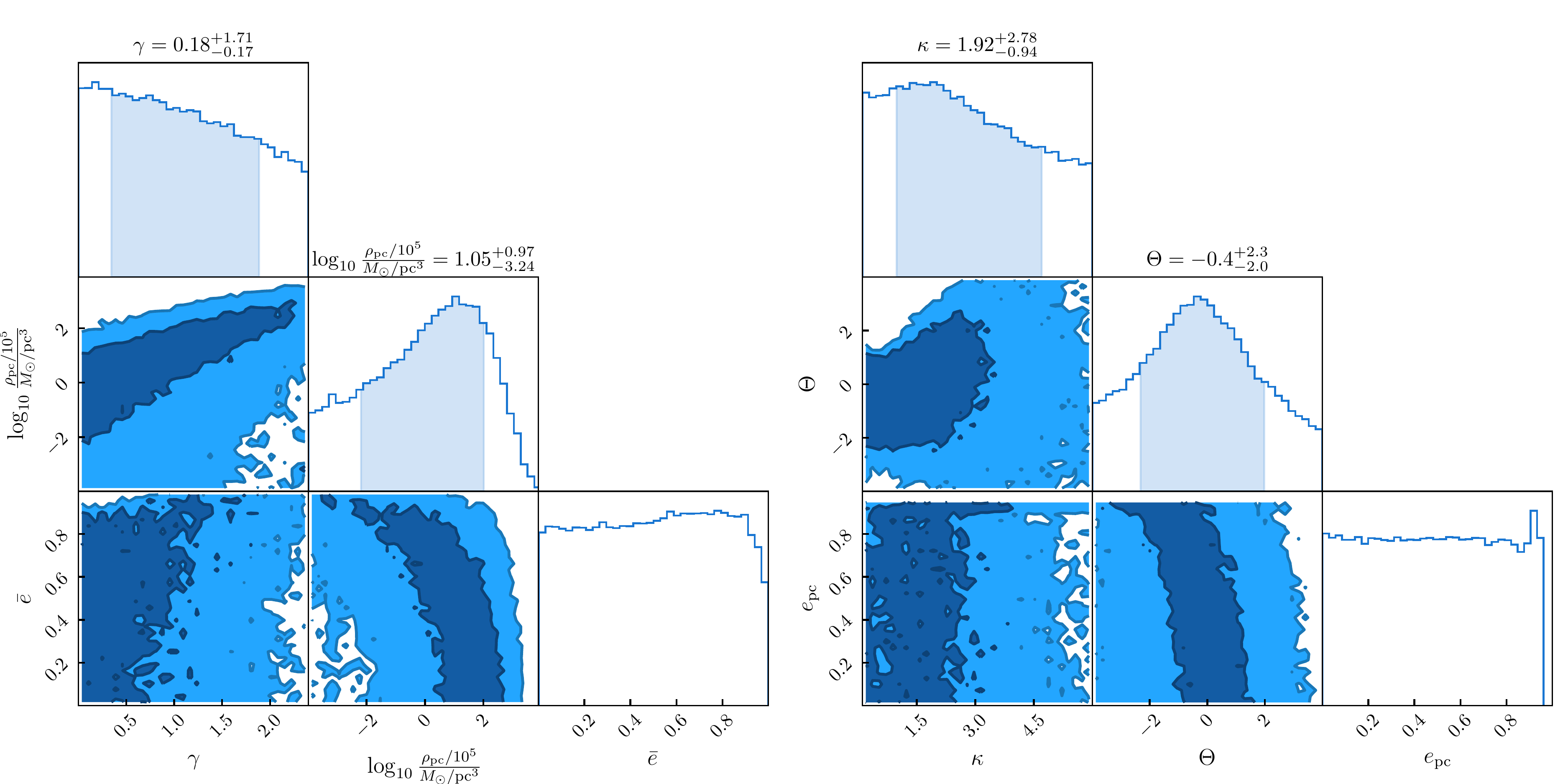}
    \caption{\textbf{Posterior constraints from alternative eccentricity and density-profile parameterizations.}
\textbf{Left:} The universal initial eccentricity $e_0$ is replaced by the average eccentricity $\bar{e}_0$ of a power-law distribution. 
\textbf{Right:} The universal density-profile parameters $(\rho_{\rm pc}, \gamma)$ are replaced by a Gaussian distribution of $\log_{10}[\tilde{\rho}_i/\tilde{\sigma}_i]$, characterized by a mean $\Theta$ and width $\kappa$, together with the eccentricity at $a = 1$\,pc, denoted $e_{\rm pc}$.}
    \label{fig:constrainte0barThetakappa}
\end{figure}

In Methods, we also present analyses that allow each SMBHB to deviate from a universal $(e_0, \rho_{\rm pc}, \gamma)$, including a power-law distribution of initial eccentricities and a variance in the initial density. The corresponding posterior constraints are shown in Fig.~\ref{fig:constrainte0barThetakappa}. These alternative parameterizations do not alter the conclusion that the posterior distribution favors $\rho_{\rm pc} \sim 10^6\,M_\odot/\text{pc}^3$. Furthermore, variations in the SMBHB population distribution merely shift the best-fit region in accordance with the scaling relations from equation~(\ref{eq:ft}).

\section*{Discussions}

In the vicinity of a binary BH, a test particle can extract orbital energy through several scatterings with each BH component. Within the PTA observation band, three-body ejections are substantially more efficient than two-body dynamical friction, especially for comparable-mass binaries, resulting in pronounced scouring of the surrounding matter distribution. In this study, we explore the potential imprints that stars or particle dark matter surrounding SMBHBs could leave on the spectrum of the SGWB. Using the NANOGrav 15-year observational data, we constrain the relevant parameter space. Our results support the occurrence of three-body scatterings with a reasonable density distribution before scouring, primarily inferred from the low-frequency turnover in the SGWB spectrum, assuming no other dominant environmental influences are present. 
Continued monitoring with existing PTAs will lead to improved constraints on the low-frequency turnover. Future observations from FAST~\cite{Xu:2023wog}, DSA-2000~\cite{Hallinan:2019qyo}, the Square Kilometre Array~\cite{Weltman:2018zrl}, and next-generation astrometry missions~\cite{10.3389/fspas.2018.00011} are expected to deliver substantially more precise spectral measurements~\cite{Caliskan:2023cqm}, promising to shed light on the physical origin of the turnover.

Resolving the degeneracy with the initial eccentricity is essential if we are to precisely determine the density profile. Potential strategies for determining the SMBHB eccentricity include resolving individual binaries through either GW or electromagnetic observations~\cite{Ayzenberg:2023hfw,DOrazioCharisi:2023} and examining correlations among different frequency bins. This latter method is based on the observation that high eccentricity contributes to several integer multiples of frequency bins simultaneously~\cite{Raidal:2024odr}. Nevertheless, setting a stringent upper limit on the galactic-centre density distribution is feasible, as both other environmental influences and nonzero orbital eccentricity tend to further elevate the turnover frequency.

The three-body slingshot mechanism considered in this study assumes that the test particles interact purely gravitationally. It, thus, applies to both stars and cold, collisionless dark matter. For a star-dominant distribution, such findings could reveal the star formation and relaxation rates near SMBHs in the galactic centre and may indicate a density higher than previously expected~\cite{Kelley:2016gse,Chen:2016zyo}.
If dark matter is the dominant density, the findings here could provide crucial insights into two longstanding questions: the identification of the particle nature of dark matter and the measurement of its density at galactic centres. Further studies are needed to evaluate how specific dark matter models, especially those that depart from the cold dark matter scenario, influence the applicability of three-body ejection. In Supplemental Information, we also discuss possible non-cold dark matter effects and other environmental mechanisms that may contribute to the observed turnover.

\section*{Methods}

\noindent{\bf Evolution of Eccentric Orbits}

We investigate the orbital evolution of SMBHBs driven by both three-body ejection of stars~\cite{Quinlan:1996vp} and cold dark matter~\cite{Milosavljevic:2001vi}, and GW emission. Our analysis incorporates eccentric orbits with evolving eccentricity~\cite{Enoki:2006kj,Chen:2016zyo,Raidal:2024odr}. Other possible environmental effects~\cite{Gould:1999ia,Armitage:2002uu,Sesana:2010qb,Sampson:2015ada,Kelley:2017lek,Bortolas:2021say,Aghaie:2023lan,Alonso-Alvarez:2024gdz,Dutra:2024qac,Ikeda:2020xvt,Broadhurst:2023tus,Koo:2023gfm,Bromley:2023yfi,Aurrekoetxea:2023jwk,Aurrekoetxea:2024cqd,Guo:2025pea,Tiede:2023dwq,Dittmann:2023dss,Ghoshal:2023fhh,Shen:2023pan,Hu:2023oiu} are discussed in Supplemental Information.

The coupled equations that govern the semimajor axis $a$ and eccentricity $e$ are as follows:
\be\begin{split}
\frac{\d a(e,a)}{\d t}=&  -\frac{64}{5} \frac{G^3 M^3}{c^5a^3} \frac{q}{(1 +q)^2} \frac{(1+\frac{73}{24} e^2 + \frac{37}{96} e^4)}{(1-e^2)^{7/2}} -H G\frac{\rho_i}{\sigma_i} a^2,\\
\frac{\d e(e,a)}{\d t} =& -\frac{304}{15} \frac{G^3 M^3}{c^5 a^4} \frac{q}{(1+q)^2} \frac{e(1+\frac{121}{304} e^2 )}{(1-e^2)^{5/2}} + H K(e,a) G\frac{\rho_i}{\sigma_i} a.
\label{eq:hardenGWe}
\end{split}\ee 
The first terms on the right-hand side of these equations correspond to GW emission~\cite{Peters:1963ux}, and the second terms describe the effects of three-body ejections~\cite{Quinlan:1996vp}. The dimensionless parameters $H$ and $K$ represent the hardening and eccentricity growth rates, respectively, as derived from scattering simulations. For hard binaries, $H$ values typically range from $15$ to $20$~\cite{Quinlan:1996vp}; in this study, we have fixed $H$ at $18$. The function for $K$ is approximated by~\cite{Quinlan:1996vp,Sesana:2006xw,2015ApJ...810...49V,Fastidio:2024crh}:
\be K(e,a) \approx
0.3 \, e\left(1-e^2\right)^{0.6}\left(1+\frac{a}{0.2\,a_h}\right)^{-1},\ee
which becomes most effective when $a$ is within the hardening radius $a_h \equiv r_i q/(4(1+q)^2)$.

Each evolutionary case begins at the influence radius, $r_i$, determined by solving the relation $\textstyle\int_0^{r_i}\! \rho\, 4\pi r^2 dr = 2M$~\cite{Frank:1976uy} for a power-law distribution $\rho(r) = \rho_{\rm pc} (r/1\,\text{pc})^{-\gamma}$ before scouring. Consequently, $\rho_i \equiv \rho(r_i)$ and $\sigma_i \equiv \sigma(r_i) = \sqrt{GM/r_i}$ are calculated. The initial eccentricity, $e_0$, defined at $r_i$, serves as a fitting parameter to be constrained.

We vary both the initial eccentricity, $e_0$, and the density profile parameters $(\rho_{\rm pc}, \gamma)$. We then calculate the orbital evolution for the SMBHB with parameters $(M, q)$ from the determined $r_i$. The SGWB spectrum is subsequently calculated by integrating over the SMBHB population parameters $(M, q, z)$, considering the density distribution $\mathrm{d}^3 \eta/(\mathrm{d}z \, \mathrm{d}M \, \mathrm{d}q)$, as referenced in~\cite{Phinney:2001di,Enoki:2006kj,Huerta:2015pva}:
\be
h_c^2(f)=\frac{4 G}{c^2 \pi f} \int \d z \d M \d q \frac{\d^3 \eta}{\d z \d M \d q} \frac{\d E_{\mathrm{GW}}}{\d f_s}\Big|_{f_s=(1+z)f}.
\ee
In this calculation, we neglect Poisson fluctuations in the SMBHB distribution, as these have a minor impact on the lowest frequency bins of the NANOGrav $15$-year dataset~\cite{Sesana:2008mz,NANOGrav:2023pdq,Lamb:2024gbh,Sato-Polito:2024lew,Xue:2024qtx}.

The GW emission spectrum, $\mathrm{d}E_{\mathrm{GW}}/\mathrm{d} \ln f_s$, calculated at the source frame frequency $f_s$, includes contributions from various orbital frequencies $f_{\rm orb}^n = f_s/n$ for integer $n > 0$. It is expressed as~\cite{Peters:1963ux}:
\be 
\frac{\d E_{\mathrm{GW}}}{\d f_s}\Big|_{f_s=(1+z)f} = \sum_{n=1}^{+\infty} \frac{\d E_{\mathrm{GW}}^n/\d t}{ n\d f_{\rm orb}^n/\d t}\Big|_{f_{\rm orb}^n= (1+z)f/n},\label{eq:summation} 
\ee
where
\begin{equation}
    \begin{aligned}
        \frac{\d E_{\mathrm{GW}}^n}{ \d t} &= \frac{32G^4M^5}{5 c^5a^5} \frac{q^2}{(1+q)^4} g(n,e),\\
        \frac{\d f_{\rm orb}^n}{\d t} &= -\frac{3\sqrt{G M}}{4\pi a^{5/2}}\frac{\d a}{\d t}.
    \end{aligned}.
\end{equation}
The orbital frequency is related to $a$ via Kepler's law $a^3 = GM/(2\pi f_{\rm orb}^n)^2$. The function $g(n, e)$, defined as
\be
\begin{aligned}
g(n, e) & \equiv \frac{n^4}{32}\bigg{[} \Big{\{}J_{n-2}(n e)-2 e J_{n-1}(n e) +\frac{2}{n} J_n(n e)+2 e J_{n+1}(n e)-J_{n+2}(n e) \Big{\}}^2 \\
& +\left(1-e^2\right)
\Big{\{}J_{n-2}(n e)-2 J_n(n e)+J_{n+2}(n e)\Big{\}}^2 +\frac{4}{3 n^2} J_n^2(n e)\bigg{]},
\end{aligned}\ee
converges to $g(2, 0) = 1$ for circular orbits. Here, $J_n$ denotes the Bessel function of the first kind of order $n$.

In practice, we employ several methods to enhance the efficiency of numerical computation. In equation~(\ref{eq:summation}), we apply a cutoff to the summation over $n$, neglecting all contributions for $n > n_{\rm max}$, with $n_{\rm max} =  3 n_{\rm peak}(e_{\rm max})$. The value of $n_{\rm peak}(e)$ is given by~\cite{Hamers:2021eir}:
\begin{equation}
    n_{\rm peak}(e) \approx 2\left(1+\sum_{k=1}^4 c_k e^k\right)\, (1-e^2)^{-3/2},
\end{equation}
where $c_1=-1.01678$, $c_2=5.57372$, $c_3=-4.9271$, and $c_4=1.68506$ and $e_{\rm max}$ represents the maximum eccentricity throughout the orbital evolution. Additionally, when $n_{\rm max}$ exceeds $100$, we sum the contributions for $n > 101$ using logarithmic steps (i.e., $n = 101, 102, 104, 108, \dots$) and implement numerical integration over $n$ using \texttt{numpy.trapz}.

\noindent{\bf Statistics}

In our data analysis, we estimate the posterior distribution of $\log_{10}\rho_{\rm pc}$, $\gamma$, $e_0$ and $\log_{10}N$. This can be separated into two components:
\begin{equation}
    \begin{aligned}
         &P\left(\log_{10}\frac{\rho_{\rm pc}/10^5}{M_{\odot}/{\rm pc}^3},\gamma,e_0,\log_{10}N|\delta t\right) \\
         =  &\int \prod_k\d\log_{10}h_{c}^2(f_k)\,P\left(\log_{10}\frac{\rho_{\rm pc}/10^5}{M_{\odot}/{\rm pc}^3},\gamma,e_0,\log_{10}N|\{\log_{10}h_{c}^2(f_k)\}\right)\times P\left(
         \{\log_{10}h_{c}^2(f_k)\}|\delta t
         \right),\label{eq:posterior_expansion}
    \end{aligned}
\end{equation}
where $\delta t$ represents the timing residual data from the NANOGrav 15-year dataset~\cite{NANOGrav:2023gor}, and $f_k = k/T_{\rm obs}$ denotes the frequencies indexed by $k=1,\,2,\,3,\,4,\,5$, corresponding to the five lowest frequency bins, with $T_{\rm obs} \approx 16$ years as the observation time.

The second term in equation~(\ref{eq:posterior_expansion}) represents the posterior distributions of the free spectrum derived by the NANOGrav collaboration~\cite{NANOGrav:2023hfp}. The first term is calculated using Bayes' theorem:
\begin{equation}
    \begin{aligned}
        &P\left(\log_{10}\frac{\rho_{\rm pc}/10^5}{M_{\odot}/{\rm pc}^3},\gamma,e_0,\log_{10}N|\{\log_{10}h_{c}^2(f_k)\}\right) \\
        \propto& P\left(\{\log_{10}h_{c}^2(f_k)\}|\log_{10}\frac{\rho_{\rm pc}/10^5}{M_{\odot}/{\rm pc}^3},\gamma,e_0,\log_{10}N\right)P\left(\log_{10}\frac{\rho_{\rm pc}/10^5}{M_{\odot}/{\rm pc}^3}\right)P\left(\gamma\right)P\left(e_0\right)P\left(\log_{10}N\right),
    \end{aligned}
\end{equation}
The first term is computed in section~{\bf `Evolution of Eccentric Orbits'}, where Poisson fluctuations are neglected. The remaining terms are the priors for each parameter:
\begin{equation}
\begin{aligned}
    &P\left(\log_{10}\frac{\rho_{\rm pc}/10^5}{M_{\odot}/{\rm pc}^3}\right) = U[-4,4],\\
    &P\left(\gamma\right) = U[0,2.4],\\
    &P\left(e_0\right) = U[0,0.999],\\
    &P\left(\log_{10}N\right) = \mathcal{N}(-1.56,1.12).
\end{aligned}
\end{equation}
where $U$ denotes a uniform distribution, $\mathcal{N}$ denotes a normal distribution and the prior for $N$ is calculated in Supplementary Information.

For practical implementation, we use \texttt{PTArcade}~\cite{Mitridate:2023oar}, which is based on \texttt{ceffyl}~\cite{Lamb:2023jls}, to model the prior distributions and fit the results to the free spectrum data.

\noindent{\bf Alternative Parameterization}

 In the main text, we employed a universal parameterization for the initial eccentricity and density profile before scouring, $(e_0, \rho_{\rm pc}, \gamma)$, for each SMBHB. In reality, each of these parameters can have its own distribution. In this section, we consider alternative parameterizations that allow for a distribution of initial eccentricities $e_0$ and a variance in $\rho_i/\sigma_i$ across the SMBHB population. We show that these alternative choices have only a minor impact on our conclusion that the parsec-scale density is approximately $10^6\,M_\odot/\textrm{pc}^3$.

First, we consider a power-law eccentricity distribution~\cite{Moe_2017}:
\begin{equation}
f(e_0) = \xi\,e_0^{\xi - 1},
\label{e0PL}
\end{equation}
where $\xi > 0$ is an exponent controlling the form of the distribution, and the distribution is normalized to unity. For example, $\xi = 2$ corresponds to the thermal distribution~\cite{1919MNRAS..79..408J}, $\xi = 1$ yields a uniform distribution, and $\xi \rightarrow 0$ strongly suppresses high eccentricities. In the left panel of Fig.~\ref{fig:constrainte0barThetakappa}, we show the posterior distribution obtained by replacing the universal $e_0$ in Fig.~\ref{fig:constraint} with the average value from this power-law distribution, $\bar{e}_0 = \xi/(\xi+1)$, while leaving the other parameters unchanged. We find that the resulting posterior distribution differs only slightly from the universal-parameterization case shown in Fig.~\ref{fig:constraint}.

We next consider alternatives to the density-profile parameters. We employ a Gaussian distribution for
\be \log_{10} [\tilde{\rho}_i/\tilde{\sigma}_i] \in \mathcal{N} (\Theta, \kappa),\label{eq:Thetakappa}\ee
with mean value $\Theta$ and Gaussian width $\kappa$, where
$\tilde{\rho}_i \equiv \rho_i/(10^5\,M_\odot/\textrm{pc}^3)$ and $\tilde{\sigma}_i \equiv \sigma_i/(10^{-3}\,c)$. Accordingly, the initial eccentricity is replaced by the eccentricity defined at $a = 1$\,pc, $e_{\rm pc}$. The corresponding posterior distribution is shown in the right panel of Fig.~\ref{fig:constrainte0barThetakappa}.

Note that this distribution can exhibit high eccentricities $e_{\rm pc}$ within the $1\,\sigma$ region even in the absence of surrounding matter for three-body ejection. This arises from the difference in definition between $e_{\rm pc}$ and $e_0$ in the main text: $e_0$ is defined at the influence radius $r_i$, which is typically much larger than $1$~pc for low-density environments. As a result, $e_{\rm pc}$ corresponds to a more extreme value of $e_0$ at low densities.

\bmhead{Acknowledgements}

We are grateful to Kfir Blum, Vitor Cardoso, Gregorio Carullo, Siyuan Chen, James Cline, Hyungjin Kim, Bin Liu, Yiqiu Ma, Zhen Pan, Shreyas Tiruvaskar, and Rodrigo Vicente for useful discussions. 

The NANOGrav Collaboration receives support from National Science Foundation (NSF) Physics Frontiers Center (Award Nos. 1430284 and 2020265), the Gordon and Betty Moore Foundation, NSF AccelNet (Award No. 2114721), an NSERC Discovery Grant and CIFAR. The Arecibo Observatory is a facility of the NSF operated under a cooperative agreement (Agreement No. AST-1744119) by the University of Central Florida in alliance with Universidad Ana G. Méndez and Yang Enterprises, Inc. The Green Bank Observatory is a facility of the NSF operated under cooperative agreement by Associated Universities, Inc. The National Radio Astronomy Observatory is a facility of the NSF operated under cooperative agreement by Associated Universities, Inc. Part of this research was performed at the Jet Propulsion Laboratory, under contract with NASA. 

Y.C. is supported by the Villum Foundation (Grant No. 37766), by the Danish Research Foundation and by the European Union's H2020 ERC Advanced Grant `Black holes: gravitational engines of discovery' (Grant Agreement No. Gravitas-101052587). The views and opinions expressed here are, however, those of the authors only and do not necessarily reflect those of the European Union or the European Research Council. Neither the European Union nor the granting authority can be held responsible for them. D.J.D. acknowledges support from the Danish Independent Research Fund through a Sapere Aude Starting Grant (No. 121587). A. Mitridate and X.X. are supported by the Deutsche Forschungsgemeinschaft under Germany's Excellence Strategy (EXC 2121 Quantum Universe – 390833306). IFAE is partially funded by the CERCA programme of the Generalitat de Catalunya. X.X. is funded by Grant No. CNS2023-143767 through MICIU/AEI/10.13039/501100011033 and by European Union NextGenerationEU/PRTR. Y.C. and X.X. acknowledge the support of the Rosenfeld Foundation and the European Consortium for Astroparticle Theory in the form of an exchange travel grant. 

L.B. acknowledges support from the NSF (Award No. AST-1909933) and from the Research Corporation for Science Advancement under Cottrell Scholar Award No. 27553. P.R.B. is supported by the Science and Technology Facilities Council (Grant No. ST/W000946/1). S.B.-S. gratefully acknowledges the support of a Sloan Fellowship, and the support of the NSF (Award No. 1815664). The work of R.B., N. Laal, X.S., J.T. and D.W. is partly supported by the George and Hannah Bolinger Memorial Fund managed by the College of Science at Oregon State University. M.C., P.P. and S.R.T. acknowledge support from the NSF (Award No. AST-2007993). M.C. was supported by the Vanderbilt Initiative in Data Intensive Astrophysics Fellowship. Support for this work was provided by the NSF through the Grote Reber Fellowship Program administered by Associated Universities, Inc./National Radio Astronomy Observatory. Pulsar research at UBC is supported by an NSERC Discovery Grant and by CIFAR. K.C. is supported by a UBC Four Year Fellowship (6456). M.E.D. acknowledges support from the Naval Research Laboratory through NASA (Contract No. S-15633Y). T.D. and M.T.L. are supported by an NSF Astronomy and Astrophysics Grant (Award No. 2009468). E.C.F. is supported by NASA (Award No. 80GSFC24M0006). G.E.F., S.C.S. and S.J.V. are supported by the NSF (Award No. PHY-2011772). K.A.G. and S.R.T. acknowledge support from an NSF CAREER award (Award No. 2146016). A.D.J. and M.V. acknowledge support from the Caltech and Jet Propulsion Laboratory President's and Director's Research and Development Fund. A.D.J. acknowledges support from the Sloan Foundation. N. Laal acknowledges support from a Larry W. Martin and Joyce B. O'Neill Endowed Fellowship in the College of Science at Oregon State University. Part of this research was carried out at the Jet Propulsion Laboratory, California Institute of Technology, under a contract with NASA (Contract No. 80NM0018D0004). D.R.L. and M.A.M. are supported by the NSF (Grant No. 1458952). M.A.M. is supported by the NSF (Grant No. 2009425). C.M.F.M. was supported in part by the NSF (Grant Nos. NSF PHY-1748958 and AST-2106552). The Dunlap Institute is funded by an endowment established by the David Dunlap family and the University of Toronto. K.D.O. was supported in part by the NSF (Grant No. 2207267). T.T.P. acknowledges support from the Extragalactic Astrophysics Research Group at Eötvös Loránd University, funded by the Eötvös Loránd Research Network, which was used during the development of this research. H.A.R. is supported by NSF Partnerships for Research and Education in Physics (Award No. 2216793). S.M.R. and I.H.S. are CIFAR fellows. Portions of this work performed at NRL were supported by ONR 6.1 basic research funding. J.D.R. also acknowledges support from start-up funds from Texas Tech University. J.S. is supported by an NSF Astronomy and Astrophysics Postdoctoral Fellowship (Award No. AST-2202388) and acknowledges previous support from the NSF (Award No. 1847938). C.U. acknowledges support from BGU (Kreitman fellowship), and the Council for Higher Education and Israel Academy of Sciences and Humanities (Excellence fellowship). C.A.W. acknowledges support from CIERA, the Adler Planetarium and the Brinson Foundation through a CIERA-Adler postdoctoral fellowship. O.Y. is supported by a NSF Graduate Research Fellowship (Grant No. DGE-2139292).
\\
\bmhead{Author Contributions Statement} 
This paper uses a decade's worth of pulsar-timing observations and is the product of the work of many people. 

Y.C., L.S. and X.X. initiated the project and developed the core idea, with D.J.D. contributing to its development. Y.C., M.D., D.J.D., X.F., A. Mitridate, L.S., and X.X. participated in discussions, provided critical feedback, and shaped the research and analysis. Y.C. coordinated the project and wrote the paper. M.D., X.F. and X.X. developed the analysis code, created the figures and edited the text. X.F. performed the analysis using alternative parameterizations, under the guidance of Y.C. and X.X. 

D.J.D. offered guidance on the SMBHB population model and the \texttt{holodeck} code, wrote the discussion on the environmental effects from gas and edited the text. A. Mitridate provided guidance on the \texttt{PTArcade} code and the presentation of NANOGrav 15-year data.

G.A., A.A, A.M.A., Z.A., P.T.B., P.R.B., H.T.C.,
K.C., M.E.D, P.B.D., T.D., E.C.F, W.F., E.F., G.E.F.,
N.G.D., D.C.G., P.A.G., J.G., R.J.J., M.L.J., D.L.K.,
M.K., M.T.L., D.R.L., J.L., R.S.L., A. McEwen, M.A.M.,
N.M., B.W.M., C.N., D.J.N., B.B.P.P., N.S.P.,
H.A.R., S.M.R., P.S.R., A.S., C.S., B.J.S.A., I.H.S.,
K.S., A.S., J.K.S., and H.M.W. developed timing models and ran observations for the NANOGrav 15 yr data
set.
Development of the \texttt{holodeck} population modeling framework was led by L.Z.K., with contributions from J.A.C-C., D.W., E.C.G., K.G., M.S.S., and S.C. 
\texttt{PTArcade}, which was used in this analysis, was mainly developed by A. Mitridate., with help from D.W., K.D.O. G.A., A.A., A.M.A., Z.A., J.G.B., P.T.B., B.B., L.B., A.B., P.R.B., S.B.-S., R.B., J.A.C.-C., M.C., S.C., T.C., J.M.C., N.J.C., F.C., H.T.C., K.C., M.E.D., P.B.D., H.D., L.D., T.D., E.C.F., W.F., E.F., G.E.F., N.G.-D., P.A.G., K.A.G., J.G., D.C.G., K.G., J.S.H., R.J.J., A.D.J., M.L.J., L.Z.K., M.K., J.S.K., N. Laal, M.T.L., W.G.L., B.L., T.J.W.L., N. Lewandowska, T.L., D.R.L., J.L., R.S.L., C.-P.M., D.R.M., A. McEwen, J.W.M., M.A.M., N.M., B.W.M., P.M.M., C.M.F.M., A. Mitridate, C.N., D.J.N., S.K.O., K.D.O., T.T.P., B.B.P.P., P.P., N.S.P., H.A.R., S.M.R., P.S.R., J.D.R., J.C.R., A. Saffer, S.C.S., A. Schmiedekamp, C.S., K. Schmitz, B.J.S.-A., X.S., J.S., M.S.S., S.V.S.F., I.H.S., D.R.S., K. Stovall, A. Susobhanan, J.K.S., J.T., S.R.T., J.E.T., C.U., M.V., R.V.H., S.J.V., H.M.W., C.A.W., D.W. and O.Y. developed and validated the NANOGrav 15-year data.

\bmhead{Data Availability Statement} 
The 15-year NANOGrav dataset is described in and available through Ref.~\cite{NANOGrav:2023hfp}.

\bmhead{Code Availability Statement} 
All methods required to evaluate the conclusions in this paper are provided in the main text, Methods and Supplementary Information. 
The \texttt{holodeck} and \texttt{PTArcade} codes used in this work are publicly available at \url{https://github.com/nanograv/holodeck} and \url{https://github.com/andrea-mitridate/PTArcade}, respectively. The code supporting the findings of this study is available via GitHub
at \url{https://github.com/XueXiao-Physics/NG15_Galactic_Tomography}.

\bmhead{Competing Interests Statement} 
The authors declare no competing interests.\\

\clearpage
\renewcommand{\figurename}{Supplementary Fig}
\setcounter{figure}{0}
\setcounter{table}{0}

\begin{center}
{\Large \bfseries Supplementary Information}\\[0.6em]
{\Large \bfseries ``Inference on inner galaxy structure via gravitational waves from supermassive binaries''}\\[0.6em]
\end{center}

\section*{Supermassive Black Hole Binary Population}

This section details the semi-analytic SMBHB population model utilized in this study, characterized by the comoving volumetric number density of SMBHBs, $\mathrm{d}^3 \eta/(\mathrm{d}z \, \mathrm{d}M \, \mathrm{d}q)$, alongside astrophysical priors on its normalization. To generate the initial SMBHB population before binary evolution, we use the \texttt{holodeck} code, following NANOGrav's astrophysical interpretation paper~\cite{NANOGrav:2023hfp}. However, we replaced the phenomenological environmental prescription~\cite{Sampson:2015ada,NANOGrav:2023hfp,Ellis:2023dgf,Saeedzadeh:2023biq,Sah:2024oyg} with the three-body scattering prescription detailed in section~{\bf Evolution of Eccentric Orbits}.

The semi-analytic model is predicated on galaxy merger rates, the SMBH-host galaxy relationship, and cosmological expansion, and is parameterized as follows~\cite{Sesana:2013wja,Sesana:2016yky}:
\be 
\frac{\mathrm{d}^3 \eta}{\mathrm{d} z \, \mathrm{d} M \, \mathrm{d} q}=\frac{\mathrm{d}^3 \eta_{\text{gal-gal}}}{\mathrm{d} z \, \mathrm{d} m_{\star1} \, \mathrm{d} q_{\star}} \frac{\mathrm{d} m_{\star1}}{\mathrm{d} M} \frac{\mathrm{d} q_{\star}}{\mathrm{d} q}, \ee
where ${\mathrm{d}^3 \eta_{\text{gal-gal}}}/{(\mathrm{d} z \, \mathrm{d} m_{\star1} \, \mathrm{d} q_{\star})}$ represents the volumetric galaxy merger rate density as a function of redshift $z$, the stellar mass of the primary galaxy $m_{\star1}$, and the mass ratio of the galaxies $q_\star \leq 1$. The galaxy merger rate can be expressed as~\cite{Chen:2018znx}:
\be
\frac{\mathrm{d}^3 \eta_{\text{gal-gal}}}{\mathrm{d} z \, \mathrm{d} m_{\star1} \, \mathrm{d} q_{\star}}  = \frac{\Psi(m_{\star1}, z')}{m_{\star1} \ln(10)} \frac{P(m_{\star1}, q_\star, z')}{T_{\text{gal-gal}}(m_{\star1}, q_\star, z')} \frac{\d t}{\d z'}.
\ee
Here, $\Psi(m_{\star1}, z')$ represents the galaxy stellar mass function (GSMF), $P(m_{\star1}, q_\star, z')$ denotes the galaxy pair fraction (GPF), and $T_{\text{gal-gal}}(m_{\star1}, q_\star, z')$ refers to the galaxy merger timescale (GMT). The variable $z'$ is defined as the advanced redshift satisfying $t(z)-t(z')= T_{\text{gal-gal}}(z')$, with the current epoch at $t(0) = 13.79$\,Gyr. The rate of change in cosmic time with respect to redshift, $\d t/\d z$, is calculated as $1/((1+z)H(z))$, following the standard cosmological model where the Hubble parameter $H(z) = H_0\left[\Omega_{\Lambda} + (1+z)^3 \Omega_m\right]^{1/2}$ with $H_0 = 67.4\, \mathrm{km}\,\mathrm{s}^{-1}\,\mathrm{Mpc}^{-1}, \Omega_m = 0.315$, and $\Omega_{\Lambda} = 0.685$. The terms ${\mathrm{d} m_{\star1}}/{\mathrm{d} M}$ and ${\mathrm{d} q_{\star}}/{\mathrm{d} q}$ are derived from the SMBH-host relation.

Below, we detail the relevant parameters in the semi-analytic model:
\begin{itemize}
\item Galaxy Stellar Mass Function (GSMF) $\Psi(m_{\star1}, z)$: This function describes the distribution of primary galaxy masses $m_{\star1}$ at different redshifts $z$, following the Schechter function~\cite{Schechter:1976iz}:
    \be
    \Psi(m_{\star1}, z) = \ln(10) \Psi_0 \left( \frac{m_{\star1}}{M_\psi} \right)^{\alpha_\psi} \exp \left( - \frac{m_{\star1}}{M_\psi} \right),
    \ee
where
\begin{equation}
\begin{aligned}
\log _{10}\left(\Psi_0 / \mathrm{Mpc}^{-3}\right) & =\psi_0+\psi_z \cdot z, \\
\log _{10}\left(M_\psi / \mathrm{M}_{\odot}\right) & =m_{\psi, 0}+m_{\psi, z} \cdot z, \\
\alpha_\psi & =1+\alpha_{\psi, 0}+\alpha_{\psi, z} \cdot z.
\end{aligned}
\end{equation}

     \item Galaxy Pair Fraction (GPF) $P(m_{\star1}, q_\star, z)$: This parameterizes the relative number of observable galaxy pairs to total galaxies, following~\cite{Chen:2018znx}:
\be P(m_{\star1}, q_\star, z) = P_0 \left( \frac{m_{\star1}}{10^{11} M_\odot} \right)^{\alpha_p} (1 + z)^{\beta_p} q_\star^{\gamma_p},\ee
    where
    \begin{equation}
\begin{aligned}
\alpha_p & =\alpha_{p, 0}+\alpha_{p, z} \cdot z, \\
\gamma_p & =\gamma_{p, 0}+\gamma_{p, z} \cdot z.
\end{aligned}
\end{equation}

    \item Galaxy Merger Timescale (GMT) $T_{\text{gal-gal}}(m_{\star1}, q_\star, z)$: This provides the duration for two galaxies to merge, expressed as~\cite{Chen:2018znx}:
\be T_{\text{gal-gal}}(m_{\star1}, q_\star, z) = T_0 \left( \frac{m_{\star1}}{10^{11} M_\odot / h} \right)^{\alpha_t} (1 + z)^{\beta_t} q_\star^{\gamma_t},\ee
    where
    \begin{equation}
\begin{aligned}
\alpha_t & =\alpha_{t, 0}+\alpha_{t, z} \cdot z, \\
\gamma_t & =\gamma_{t, 0}+\gamma_{t, z} \cdot z.
\end{aligned}
\end{equation}

\item SMBH-host relation: This assumes a one-to-one correspondence between galaxy pairs and SMBHBs as per Ref.~\cite{NANOGrav:2023hfp}, expressed as
\be
\log_{10}(M_{\text{BH}}/M_\odot) = \mu + \alpha_\mu \log_{10}\left( \frac{M_{\text{bulge}}}{10^{11} M_\odot} \right) + \mathcal{N}(0, \epsilon_\mu),
\ee
where $\mathcal{N}(0, \epsilon_\mu)$ denotes a normal distribution accounting for scatter with a mean of zero and standard deviation $\epsilon_\mu$ in dex. The bulge mass is calculated as a fraction of the total galaxy stellar mass, $M_{\text{bulge}} = f_{\star,\text{bulge}} \cdot m_{\star1}$, with $f_{\star,\text{bulge}}=0.615$~\cite{Lang:2014cta,Bluck:2014ila}.
    \end{itemize}

\begin{table}[h]
\centering
\begin{tabular}{|l|c|c|c|}
\hline
\textbf{Model Component} & \textbf{Symbol} & \textbf{Fiducial Value} & \textbf{Astrophysical Priors} \\ \hline

\multirow{6}{*}{GSMF $(\Psi)$} 
& $\psi_0$ & $-2.27$ & $\mathcal{N}(-2.56, 0.4)$ \\
& $\psi_z$ & $-0.60$ & $\cdots$ \\
& $m_{\psi,0}$ & $11.15$ & $\mathcal{N}(10.9, 0.4)$ \\
& $m_{\psi,z}$ & $+0.11$ & $\cdots$ \\
& $\alpha_{\psi,0}$ & $-1.21$ & $\mathcal{N}(-1.2, 0.2)$ \\
& $\alpha_{\psi,z}$ & $-0.03$ & $\cdots$ \\ \hline

\multirow{7}{*}{GPF $(P)$} 
& $P_0$ & $+0.033$ & $\cdots$ \\
& $\alpha_{p,0}$ & $0.0$ & $\cdots$ \\
& $\alpha_{p,z}$ & $0.0$ & $\cdots$ \\
& $\beta_{p,0}$ & $+1.0$ & $\mathcal{N}(0.8, 0.4)$ \\
& $\beta_{p,z}$ & $0.0$ & $\cdots$ \\
& $\gamma_{p,0}$ & $0.0$ & $\mathcal{N}(0.5, 0.3)$ \\
& $\gamma_{p,z}$ & $0.0$ & $\cdots$ \\ \hline

\multirow{7}{*}{GMT $(T_{\text{gal-gal}})$} 
& $T_0$ & $+0.5$ Gyr & $U(0.2, 5.0)$ Gyr \\
& $\alpha_{t,0}$ & $0.0$ & $\cdots$ \\
& $\alpha_{t,z}$ & $0.0$ & $\cdots$ \\
& $\beta_{t,0}$ & $-0.5$ & $U(-2.0, 0.0)$ \\
& $\beta_{t,z}$ & $0.0$ & $\cdots$ \\
& $\gamma_{t,0}$ & $-1.0$ & $\cdots$ \\
& $\gamma_{t,z}$ & $0.0$ & $\cdots$ \\ \hline

\multirow{4}{*}{$M_{\text{BH}}$-$M_{\text{bulge}}$ ($M_{\rm BH}$)} 
& $\mu$ & $8.65$ & $\mathcal{N}(8.6, 0.2)$ \\
& $\alpha_{\mu}$ & $+1.10$ & $\mathcal{N}(1.2, 0.2)$ \\
& $\epsilon_{\mu}$ & $0.32$ & $\mathcal{N}(0.32, 0.15)$ dex \\
& $f_{\star,\text{bulge}}$ & $+0.615$ & $\cdots$ \\ \hline

\end{tabular}
\caption{\textbf{Fiducial choices and priors used in the semi-analytic model.}
Fiducial values and astrophysical priors for the parameters of the semi-analytic model, following Ref.~\cite{NANOGrav:2023hfp}. The fiducial values for $\psi_0$, $m_{\psi,0}$, $\mu$, and $\epsilon_\mu$ correspond to the best-fit results from the `Phenom+Astro' model. $\mathcal{N}$ indicates a normal distribution and $U$ a uniform distribution.}
\label{tab:SAMP}
\end{table}

In Supplementary Table~\ref{tab:SAMP}, we list all the relevant parameters along with their fiducial values and astrophysical priors, as outlined in Ref.~\cite{NANOGrav:2023hfp}. The fiducial values for $\psi_0$, $m_{\psi,0}$, $\mu$, and $\epsilon_\mu$ are based on the best-fit results from the `Phenom+Astro' model. In this study, the comoving volumetric number density of SMBHBs follows these fiducial values. The differential distribution of the fiducial model is shown in Supplementary Fig.~\ref{fig:detadzMq}.

For the data analysis, we allow the overall normalization of $h_c^2$ to vary as a nuisance parameter, denoted by $N$, which accounts for uncertainties in the population model. This parameter is assigned a prior based on the astrophysical priors listed in Supplementary Table~\ref{tab:SAMP} that contribute to the normalization. Specifically, for each case, we calculate the corresponding $h_c^2$ for circular orbits driven solely by GW emission, and compare it to that of the fiducial model, defining the ratio as $N$. The distribution of $N$ is shown in Supplementary Fig.~\ref{fig:N}, which can be modeled as $\log_{10} N = \mathcal{N}(-1.56, 1.12)$.

\begin{figure*}[h]
    \centering
    \includegraphics[width=0.98\linewidth]{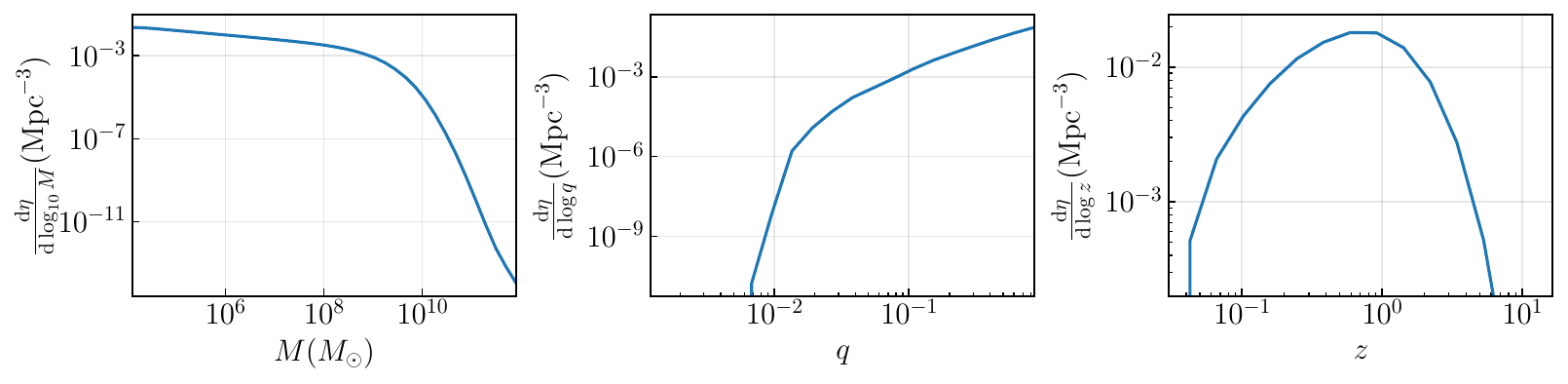}
    \caption{\textbf{Distribution of SMBHBs in mass, mass ratio, and redshift.}
Comoving volumetric number density of SMBHBs for the fiducial model, shown as a function of total SMBH mass $M$ (left), mass ratio $q$ (middle), and redshift $z$ (right).}
    \label{fig:detadzMq}
\end{figure*}

\begin{figure*}[h]
    \centering
    \includegraphics[width=0.5\linewidth]{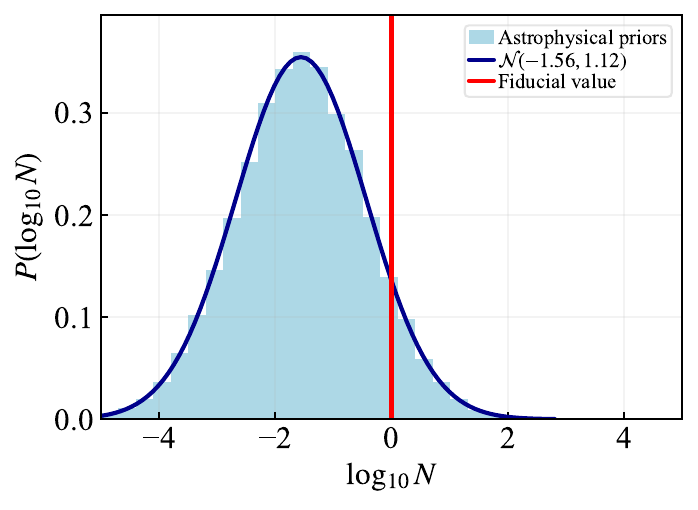}
    \caption{\textbf{Normalization distribution for SMBHB population models.}
Distribution of the normalization factor $N$, defined as the ratio between the $h_c^2$ amplitude predicted by SMBHB population models (using astrophysical priors) and the fiducial value in Supplementary Table~\ref{tab:SAMP}, evaluated without environmental hardening or eccentricity effects. The distribution is fitted as $\log_{10} N = \mathcal{N}(-1.56, 1.12)$ and adopted as the prior on the normalization in the data analysis. The fiducial value, fixed at $N=1$, is indicated by the red line.}
    \label{fig:N}
\end{figure*}

\section*{Other Possible Environmental Effects}

In principle, multiple environmental mechanisms may contribute to binary hardening at large separations, including interactions with gas~\cite{Gould:1999ia,Armitage:2002uu}, stars~\cite{Quinlan:1996vp}, and dark matter~\cite{Milosavljevic:2001vi}. In general, the orbital evolution can be written as a sum of contributions from different processes, $\mathrm{d}a/\mathrm{d}t = \sum_j \mathrm{d}a/\mathrm{d}t|_j$, where $j$ indexes distinct environmental channels~\cite{Sesana:2010qb,Sampson:2015ada,Kelley:2017lek,Bortolas:2021say}. Below we discuss environmental effects beyond the three-body scattering and eccentricity-driven evolution considered in the main analysis.

Dark matter models beyond the cold, collisionless test-particle approximation may introduce distinct forms of environmental coupling. Examples include wave-like (ultralight bosonic) dark matter~\cite{Aghaie:2023lan,Tomaselli:2024ojz,Guo:2025pea} and self-interacting dark matter~\cite{Alonso-Alvarez:2024gdz,Dutra:2024qac}. In particular, wave-like dark matter exhibits qualitatively different responses depending on the binary separation and boson mass~\cite{Ikeda:2020xvt,Broadhurst:2023tus,Koo:2023gfm,Bromley:2023yfi,Aurrekoetxea:2023jwk,Aurrekoetxea:2024cqd,Guo:2025pea}. However, these effects are highly model-dependent and require additional assumptions about the nature of dark matter.

Another potential source of environmental effects is a circumbinary disk around the SMBHB. If the galactic centers hosting SMBHBs are gas-rich and the binaries are accreting at the Eddington rate, the gas-driven hardening rate (conservatively choosing large values from Ref.~\cite{Tiede:2023dwq}), could be comparable to three-body scatterings at separations below $0.1$\,pc for $M \sim 10^{10}\,M_\odot$ ($2$\,nHz). 
However, at these separations and lower the binary will decouple from the disk~\cite{Dittmann:2023dss}, thus mitigating disk-driven hardening. Furthermore, SMBHBs of this mass cannot be surrounded by gravitationally stable thin disks until closer to merger and massive self-gravitating disks must be employed~\cite{Haiman:2009te, RoedigSesana:2014, Franchini:2021uiy}.

Another commonly discussed mechanism is two-body dynamical friction~\cite{Ghoshal:2023fhh,Shen:2023pan,Aghaie:2023lan,Hu:2023oiu}. However, this effect is already encompassed within the three-body scattering experiments~\cite{Quinlan:1996vp}. At large separations, when the binary orbital velocity is below the characteristic stellar velocity dispersion, dynamical friction dominates the early inspiral. As the binary shrinks and its orbital velocity exceeds the stellar dispersion, the evolution transitions naturally into the three-body scattering regime, which becomes the primary hardening mechanism. For the SMBHB masses and mass ratios most relevant to PTA observations, binaries spend the majority of their evolution in this three-body dominated phase.

In summary, while alternative environmental effects may contribute to binary hardening, the three-body scattering of stars and cold dark matter remains the most robust and minimal physical mechanism capable of producing the observed low-frequency turnover without requiring additional model assumptions.

\clearpage
{\Large The NANOGrav Collaboration}

\vspace{2em}

\noindent
\small
Gabriella Agazie$^{12}$,
Akash Anumarlapudi$^{12}$,
Anne M. Archibald$^{13}$,
Zaven Arzoumanian$^{14}$,
Jeremy G. Baier$^{15}$,
Paul T. Baker$^{16}$,
Bence B\'{e}csy$^{15}$,
Laura Blecha$^{17}$,
Adam Brazier$^{18, 19}$,
Paul R. Brook$^{20}$
% ...

\vspace{2em}

\noindent 
\footnotesize
$^{12}$Center for Gravitation, Cosmology and Astrophysics, Department of Physics, University of Wisconsin-Milwaukee, P.O. Box 413, Milwaukee, WI 53201, USA\\
$^{13}$Newcastle University, Newcastle upon Tyne NE1 7RU, UK\\
$^{14}$X-Ray Astrophysics Laboratory, NASA Goddard Space Flight Center, Code 662, Greenbelt, MD 20771, USA\\
$^{15}$Department of Physics, Oregon State University, Corvallis, OR 97331, USA\\
$^{16}$Department of Physics and Astronomy, Widener University, One University Place, Chester, PA 19013, USA\\
$^{17}$Physics Department, University of Florida, Gainesville, FL 32611, USA\\
$^{18}$Cornell Center for Astrophysics and Planetary Science and Department of Astronomy, Cornell University, Ithaca, NY 14853, USA\\
$^{19}$Cornell Center for Advanced Computing, Cornell University, Ithaca, NY 14853, USA\\
$^{20}$Institute for Gravitational Wave Astronomy and School of Physics and Astronomy, University of Birmingham, Edgbaston, Birmingham B15 2TT, UK\\
% ...

\end{document}